\documentclass[10pt,conference]{IEEEtran}
\IEEEoverridecommandlockouts

\usepackage{cite}
\usepackage{amsmath,amssymb,amsfonts}
\usepackage{algorithmic}
\usepackage{graphicx}
\usepackage{textcomp}
\usepackage{xcolor}
\usepackage{hyperref}
\usepackage{booktabs}
\usepackage{tabularx}
\usepackage[caption=false,font=footnotesize]{subfig}
\usepackage{placeins}
\usepackage{flushend}

\newcommand{\nsf}[1]{\href{https://www.nsf.gov/awardsearch/showAward?AWD_ID=#1}{#1}}

\newcommand{\artifacturl}{https://github.com/caslab-code/qc-fidelity-and-cost}

\begin{document}

\title{Quantum Fidelity-per-Cost: A Metric for Evaluation of Quantum Computing Systems
\thanks{$^\dagger$ The work was completed while Siddarth Shinde was a paid summer researcher at Northwestern University.\par$^\ddagger$ This work was supported in part by NSF grant \nsf{2332406}.}
}

\author{
\IEEEauthorblockN{Siddarth Shinde$^\dagger$}
\IEEEauthorblockA{\textit{Electrical and Computer Engineering} \\
\textit{Northwestern University}\\
Evanston, IL, USA \\
siddarthshinde@u.northwestern.edu}
\and
\IEEEauthorblockN{Jakub Szefer}
\IEEEauthorblockA{\textit{Electrical and Computer Engineering} \\
\textit{Northwestern University}\\
Evanston, IL, USA \\
jakub.szefer@northwestern.edu}
}

\maketitle

\begin{abstract}
Cloud-accessible quantum computing has made hardware comparison not only a physics benchmark but also a practical purchasing decision. Cost-aware comparison of quantum computers remains underexplored and is difficult to do under the heterogeneous billing models offered by various cloud-based quantum computing providers. This paper makes two main contributions to enable price-aware comparison of quantum computers. First, this work presents a cross-provider measurement study of quantum circuit execution fidelity spanning 14 cloud QPU access-path entries (12 distinct physical QPUs) across four cloud access paths: Amazon Web Services (AWS) cloud, IBM Quantum Runtime (IBM) cloud, IQM Resonance (IQM) cloud, and Oxford Quantum Circuits (OQC) cloud. Second, this work proposes and analyzes a cost-aware score, Quantum Fidelity-per-Cost (QFC), which combines Kullback--Leibler (KL) divergence from an ideal output distribution, shot count, and monetary cost into one possible metric under a documented billing model. The main empirical observation from this work is that cost-aware ranking can differ from purely fidelity-based evaluation of quantum computers, and that users may select different quantum computing backends when they consider price in their selection, as opposed to selection based on fidelity alone. This work shows that the ranking is stable under reweighting of the metric, and that a device's billing model, not its hardware, governs how its score scales with shot count. Reported QFC values change as new machines come online or as providers revise their prices.

\end{abstract}


\section{Introduction}

Quantum computing is still predominantly in the Noisy Intermediate-Scale Quantum (NISQ) regime, where noise and limited circuit depth constrain practical workloads~\cite{preskill_nisq}. At the same time, cloud access has made real quantum processing units (QPUs) broadly available through managed services and vendor APIs. For many users, choosing a QPU is no longer purely a hardware-comparison problem; it is a resource-allocation problem under a finite budget.

The challenge is that cloud billing is heterogeneous. Some providers charge a fixed task fee plus a per-shot fee, while others bill by runtime~\cite{aws_braket_pricing,ibm_quantum_pricing,iqm_resonance_pricing}. These models can reorder which devices are best for the same experiment. A high-fidelity device may still deliver low value per dollar if its access model is expensive relative to alternatives.

Many established quantum metrics target hardware capability or throughput, such as Quantum Volume and circuit layer operations per second (CLOPS)~\cite{cross_qv,wack_clops}. Application-oriented benchmark suites have also emerged, including QASMBench, SupermarQ, QPack, and the QED-C application-oriented benchmarks~\cite{li_qasmbench,tomesh_supermarq,donkers_qpack,lubinski_qedc}, alongside frameworks such as QUARK~\cite{finzgar_quark} and application-centric aggregate scores such as the Q-score~\cite{martiel_qscore}. Recent surveys additionally emphasize multi-layer benchmarking across hardware, software, cloud, and applications~\cite{lorenz_systematic_benchmarking}. However, these metrics still do not directly answer budget-constrained workload placement questions under heterogeneous cloud billing.

This paper addresses this gap empirically. We report a cross-provider measurement study of a single, narrow probe workload (a two-qubit Bell-state) executed 247 times across 14 cloud QPU access-path entries. We treat a physical QPU accessed through different commercial interfaces as separate \emph{access-path entries} when pricing or runtime semantics differ. We then introduce one concrete cost-aware metric, \emph{Quantum Fidelity-per-Cost} (QFC), that combines Kullback--Leibler (KL) divergence, shot count, and monetary cost into a single number under a documented billing model. The proposed metric (QFC) is one possible metric; other variants of the metric (for example, using different divergence scores internally) are possible. The main observation from this work is that fidelity-only and cost-aware rankings can disagree.

\subsection{Scope of the Research Study}

\noindent In this work, the workload is a single two-qubit Bell-state circuit; conclusions about deeper circuits, application workloads, or intrinsic hardware quality are not supported by this data. The pricing values, runtime fields, and calibration inputs reflect a time-bounded snapshot, and reported QFC values are expected to change as new machines come online or providers revise prices. The empirical contribution of this work is a reproducible pipeline and a dataset on which QFC and alternative cost-aware scores can be computed.

\newpage
\subsection{Contributions}

\noindent The contributions of this paper are:
\begin{itemize}
    \item \textbf{Cross-Provider Measurement Data.} We collect and release a dataset of 247 Bell-state executions across 14 QPU access-path entries (12 distinct physical QPUs) at shot counts from 20 to 1000, with per-run counts, derived total variation distance (TVD) and Kullback--Leibler (KL) divergence values, and provider-reported cost inputs.
    \item \textbf{Cost Modeling.} We catalog a four-class cross-cloud pricing taxonomy and implement consistent cost modeling for the three classes present in our run set, covering AWS Braket, IBM Quantum Runtime, IQM Resonance, and Oxford Quantum Circuits (OQC).
    \item \textbf{One Concrete Cost-Aware Score (QFC).} We define QFC, a metric combining a KL-based information-quality factor, a shot-count reward, and billed monetary cost, weighted by sensitivity parameters $\alpha$, $\beta$ and $\gamma$ and evaluated at the default $\alpha=\beta=\gamma=1$.
    \item \textbf{Ranking-Disagreement Finding.} The device with the lowest observed mean TVD at 1000 shots is not the top-ranked device by QFC. We take this to be the primary empirical finding of this work, and show it persists across a range of $(\alpha,\beta,\gamma)$ and under an alternative billing assumption.
    \item \textbf{Billing-Regime Effect.} We show that a device's billing model, not its hardware, fixes how its QFC scales with shot count, so cost-aware rankings are only meaningful at a stated shot budget.
\end{itemize}

\subsection{Code and Data Availability}

\noindent The evaluation pipeline, the saved per-run measurement counts for all 247 runs, the derived TVD, KL and QFC values, and the analysis and figure scripts are available at \url{\artifacturl}. Every figure and table in this paper, including the robustness analysis of Section~\ref{sec:sensitivity}, can be regenerated from the saved counts with no provider credentials and no re-execution on hardware, so QFC and alternative cost-aware scores can be recomputed as prices and hardware calibrations evolve.

\subsection{Paper Organization}

\noindent The rest of the paper is organized as follows. Section~\ref{sec:pricing} summarizes cloud pricing models. Section~\ref{sec:cost} defines the fidelity and cost terms used by QFC. Section~\ref{sec:qfc} presents QFC results, the billing-regime effect, access-path observations, and a robustness analysis. Section~\ref{sec:related_work} discusses related work, Section~\ref{sec:discussion} provides discussion, and Section~\ref{sec:conclusion} concludes.

\section{Cloud-Based Quantum Computing Pricing}
\label{sec:pricing}

Cloud quantum services expose different billing semantics for physically similar workloads. Because QFC explicitly depends on cost, a consistent interpretation of pricing models is required before comparing devices.

We categorize pricing into four practical classes; the representative per-device prices used to parameterize our analysis are collected in Table~\ref{tab:billing_models}. \textbf{Per-task + per-shot} charges a fixed fee per submitted task plus a variable fee proportional to shots, $C=c_{\text{task}}+N c_{\text{shot}}$, and is common on AWS Braket~\cite{aws_braket_pricing}. \textbf{Runtime-based} charges in proportion to an execution time $t$ reported by the provider, $C=c_{\text{sec}}\,t$~\cite{ibm_quantum_pricing,iqm_resonance_pricing}. \textbf{Per-shot only} bills each shot with no fixed task fee, $C=N c_{\text{shot}}$. \textbf{Per-gate execution} charges on gate usage, $C=m+c_1 N_{1q} N+c_2 N_{2q} N$, sometimes with a minimum program charge~\cite{azure_quantum_pricing}; here $N_{1q}$ and $N_{2q}$ are the single- and two-qubit gate counts of the compiled circuit, $c_1$ and $c_2$ are per-gate-shot unit prices, and $m$ is a per-program baseline, following the Azure IonQ specification. Only the first three classes appear in our run set; the per-gate class is listed for completeness and is not~evaluated.

\begin{table}[t]
\centering
\footnotesize
\caption{Representative pricing terms used in this study, grouped by billing model. Prices are a snapshot collected during the study window.}
\label{tab:billing_models}
\begin{tabularx}{\columnwidth}{@{}X l r@{}}
\toprule
\textbf{QPU} & \textbf{Access path} & \textbf{Price} \\
\midrule
\multicolumn{3}{@{}l}{\textit{Per-task + per-shot} ($\$0.30$ per task)} \\
Rigetti Ankaa-3     & AWS Braket & \$0.0009/shot \\
IQM Garnet          & AWS Braket & \$0.0014/shot \\
IQM Emerald         & AWS Braket & \$0.0016/shot \\
IonQ Aria-1         & AWS Braket & \$0.03/shot \\
\addlinespace[3pt]
\multicolumn{3}{@{}l}{\textit{Per-shot only}} \\
OQC Toshiko Tokyo-1 & OQC cloud & \$0.0013/shot \\
\addlinespace[3pt]
\multicolumn{3}{@{}l}{\textit{Runtime-based}} \\
IBM (7 devices)     & IBM Runtime & \$1.60/s \\
IQM Garnet          & IQM Resonance & \$0.50/s \\
IQM Emerald         & IQM Resonance & \$0.75/s \\
IQM Sirius          & IQM Resonance & \$0.30/s \\
\bottomrule
\end{tabularx}

\vspace{3pt}
\footnotesize
\noindent Sources: AWS Braket~\cite{aws_braket_pricing}, Azure Quantum~\cite{azure_quantum_pricing}, IBM Quantum~\cite{ibm_quantum_pricing}, IQM Resonance~\cite{iqm_resonance_pricing}. IBM runtime pricing is the pay-as-you-go rate.
\end{table}

\section{Cost-Aware Fidelity Formulation}
\label{sec:cost}

A practical cloud quantum comparison should reward two properties simultaneously: (i) closeness to the target output distribution and (ii) efficiency under the provider billing model. This section defines the fidelity and cost terms used throughout the paper.

\subsection{Fidelity Measure: Total Variation Distance}

Let $P_{\mathrm{emp}}$ denote the empirical 2-qubit output distribution from a hardware run and $P_{\mathrm{ideal}}$ denote the ideal Bell-state distribution:
$P_{\mathrm{ideal}}(00)=P_{\mathrm{ideal}}(11)=0.5$ and $P_{\mathrm{ideal}}(01)=P_{\mathrm{ideal}}(10)=0$.
We compute fidelity error using total variation distance (TVD), summing over the four two-qubit outcomes $\mathcal{X}=\{00,01,10,11\}$:
\begin{equation}
\mathrm{TVD}(P_{\mathrm{emp}},P_{\mathrm{ideal}})=\frac{1}{2}\sum_{x\in\mathcal{X}}\left|P_{\mathrm{emp}}(x)-P_{\mathrm{ideal}}(x)\right|.
\label{eq:tvd}
\end{equation}

\begin{figure*}[!t]
  \centering
  \subfloat[Mean TVD vs.\ shots.\label{fig:tvd_summary_all:a}]{%
    \includegraphics[width=0.47\linewidth]{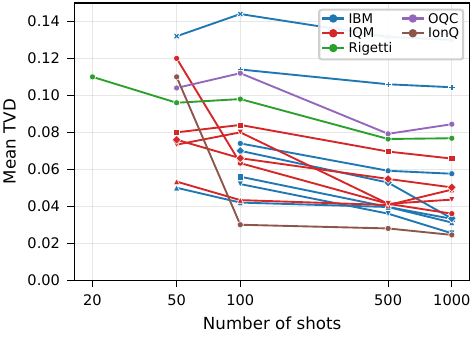}%
  }\hfill
  \subfloat[Per-device TVD at 1000 shots.\label{fig:tvd_summary_all:b}]{%
    \includegraphics[width=0.47\linewidth]{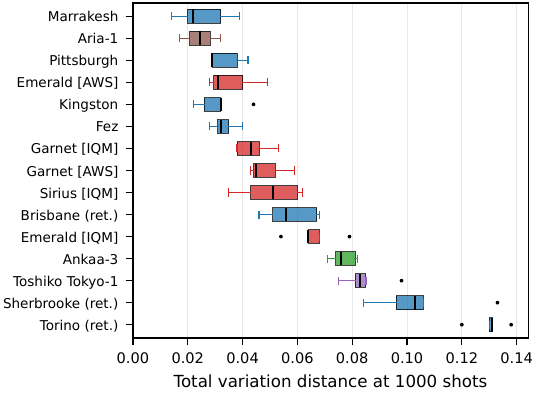}%
  }
  \caption{Bell-state fidelity error. Mean TVD falls with shot count on every provider, but the ordering of devices is largely fixed by 100 shots; at 1000 shots the spread across access paths is a factor of five, from 0.025 (IBM Marrakesh) to 0.130 (IBM Torino, retired). Color encodes provider; IQM entries are annotated with the cloud they were reached through.}
  \label{fig:tvd_summary_all}
\end{figure*}

\noindent Across the full dataset, mean TVD generally decreases as shots increase, monotonically on 8 of the 14 access paths and with small non-monotonic excursions on the rest, consistent with reduced finite-sampling noise as each device's noisy output distribution is estimated more precisely; more shots cannot remove systematic device error (Figure~\ref{fig:tvd_summary_all}).

\subsection{Cost Terms by Billing Model}

For a run with shot count $N$, total cost $C$ is computed from the provider model:
\begin{align}
C_{\text{task+shot}}(N) &= c_{\text{task}} + N\,c_{\text{shot}}, \\
C_{\text{runtime}}(t) &= c_{\text{sec}}\,t,
\end{align}
where $t$ is the provider-reported runtime (or billed duration proxy). For OQC, $c_{\text{task}}=0$ and cost is shot-only.

Runtime semantics are provider-defined and are not necessarily the same physical quantity across vendors. For IBM devices we use the \texttt{quantum\_seconds} field reported by the Qiskit Runtime service, and for IQM Resonance we use the \texttt{runtime\_seconds} field reported with the job payload. These fields may include or exclude classical overhead, queueing, or compilation time in provider-specific ways; cross-vendor runtime-based cost comparisons therefore inherit this definitional difference, and we treat it as part of the access-path semantics being evaluated rather than as a nuisance variable to be removed. The two fields also differ in resolution: \texttt{quantum\_seconds} is reported to whole seconds, so IBM cost is quantized in \$1.60 increments, while IQM's \texttt{runtime\_seconds} is reported to microseconds. For runtime-billed IQM entries we round up to the nearest whole second before applying per-second pricing, which is the conservative reading of a per-second tariff; Section~\ref{sec:sensitivity} shows that removing this rounding leaves the reported ranking~unchanged.

\subsection{QFC Metric}

To combine fidelity and cost, we use a Kullback--Leibler (KL)-based score:
\begin{equation}
\mathrm{QFC}(N)=\frac{\exp\!\left(-\alpha\,D_{\mathrm{KL}}(P_{\mathrm{emp}}\Vert P_{\mathrm{ideal}})\right)\,N^{\gamma}}{C(N)^{\beta}},
\label{eq:qfc_metric}
\end{equation}
where $\alpha,\beta,\gamma>0$ are sensitivity parameters and $C(N)$ is computed with the billing model above. A small additive smoothing constant $\epsilon=10^{-12}$ is used in KL computation to avoid undefined logarithms when ideal probabilities are zero. We use $\alpha=\beta=\gamma=1$ in most reported results below, other values of $\alpha,\beta,\gamma$ are analyzed in the sensitivity study in Section~\ref{sec:sensitivity}.

Because the ideal Bell distribution assigns zero probability to outcomes $01$ and $10$, $D_{\mathrm{KL}}(P_{\mathrm{emp}}\Vert P_{\mathrm{ideal}})$ is otherwise undefined whenever noisy runs assign nonzero probability to those outcomes. In code, KL is computed over padded 2-qubit outcomes using
\[
\sum_{x:\,P_{\mathrm{emp}}(x)>0} P_{\mathrm{emp}}(x)\,\log\!\frac{P_{\mathrm{emp}}(x)+\epsilon}{P_{\mathrm{ideal}}(x)+\epsilon},
\]
with $\epsilon=10^{-12}$ and no post-smoothing renormalization.

KL divergence is used as the fidelity term inside QFC for practical reasons rather than as a claim of unique theoretical appropriateness. The transform $\exp(-\alpha D_{\mathrm{KL}})$ maps any $D_{\mathrm{KL}}\ge 0$ into a bounded factor in $(0,1]$, where $1$ corresponds to a perfect match, so it composes with the shot-count and cost terms without a separate normalization; this is the structural property the rest of QFC relies on. KL also carries a standard information-theoretic reading as the expected log-likelihood loss incurred when coding samples from $P_{\mathrm{emp}}$ against $P_{\mathrm{ideal}}$~\cite{kullback1951}, and it is sensitive to probability mass placed on outcomes that are unlikely under the ideal target. This matches the dominant error mode observed here: noisy Bell-state runs primarily leak probability mass to the outcomes $01$ and $10$ that should have zero ideal probability.

TVD is reported throughout this paper as a complementary fidelity-error metric; QFC combines a KL-based information-quality factor with cost rather than replacing TVD. Other divergences, such as the Jensen--Shannon divergence (JSD) or the Hellinger distance, are reasonable substitutes, and we regard the choice of divergence as a specification knob rather than a uniquely determined property of QFC.

Equation~\eqref{eq:qfc_metric} creates an explicit fidelity-cost trade-off: better fidelity and larger shot counts increase QFC, while higher monetary cost decreases it. QFC is therefore a task-level, workload-specific metric under a chosen billing model and shot budget, not an intrinsic hardware-quality metric. Different divergence choices, shot weightings, or cost definitions would produce different absolute QFC values, and potentially different rankings; the purpose of the formulation here is to provide one concrete, reproducible score under which the cross-provider measurement data can be discussed.

\section{Quantum Fidelity-per-Cost Results}
\label{sec:qfc}

\subsection{Evaluation Setup}

\noindent We use the QFC definition from Equation~\eqref{eq:qfc_metric} with $\alpha=\beta=\gamma=1$ to score each run. The workload used in evaluation of QFC is a two-qubit Bell-state preparation circuit executed repeatedly on cloud QPUs. For each run we collect measurement counts, compute TVD (Equation~\eqref{eq:tvd}) and KL divergence to the ideal Bell distribution, compute run cost from the provider model, and then evaluate QFC.

Qubit selection and routing are left to each provider's default toolchain: circuits are transpiled with the vendor pass manager at its default optimization level, without a manual initial layout. This measures the access path as an ordinary user meets it, but the physical qubit pair is chosen by the provider and may differ between runs and between access paths to the same device, so we report outcomes per access path and do not attribute differences to named qubits.

\subsubsection{Vendors and QPUs Tested}

The evaluated set contains 14 QPU access-path entries over 12 distinct physical QPUs, all superconducting, and is enumerated in Table~\ref{tab:qfc_1000_rank}. IQM Garnet and IQM Emerald each appear twice because they are reached through two distinct access paths, AWS Braket and IQM Resonance.

\subsubsection{Cloud Access Paths Evaluated}

The 247 runs divide as 49 on AWS Braket, 112 on IBM Quantum Runtime, 66 on IQM Resonance and 20 on OQC Cloud. Shot levels are 50, 100, 500 and 1000 throughout, except that 20-shot data exists only for Rigetti Ankaa-3 and 50-shot data is missing for five of the seven IBM devices. Because coverage is uneven, cross-device comparisons use the largest common shot range.

\begin{figure}[!t]
    \centering
    \includegraphics[width=\linewidth]{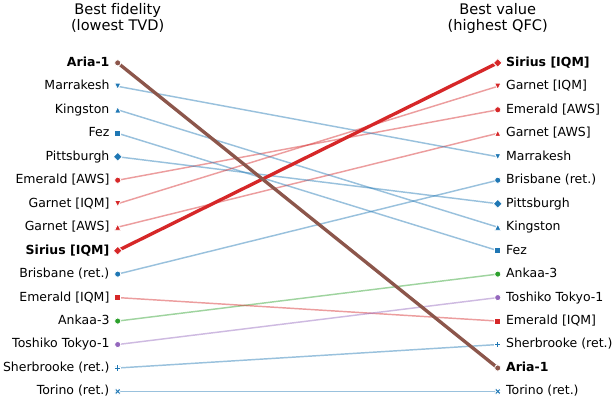}
    \caption{The 1000-shot ordering by fidelity (left) against the ordering by QFC (right). Every crossing line is an access path whose standing changes once price is accounted for. The two emphasized lines are the largest movers: trapped-ion IonQ Aria-1 has the best fidelity in the study but falls to 14th by value, while IQM Sirius rises from 9th to 1st.}
    \label{fig:rank_disagreement}
\end{figure}

\begin{figure}[!t]
    \centering
    \includegraphics[width=.95\linewidth]{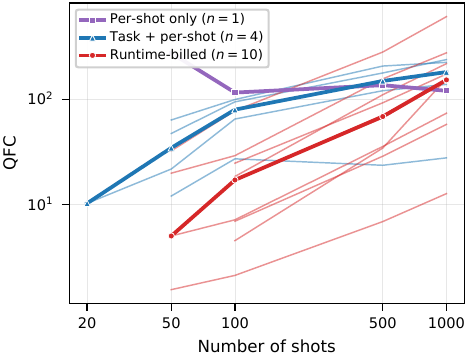}
    \caption{QFC vs.\ shot count on log--log axes ($\alpha=\beta=\gamma=1$). Thin lines are individual access paths, colored by billing model; bold lines are the median of each group. The three models produce three distinct shapes: per-shot-only billing is flat in $N$, task-plus-shot billing rises and saturates, and runtime billing rises with unit slope because the billed duration barely depends on $N$.}
    \label{fig:qfc_all_qpus}
\end{figure}

Figure~\ref{fig:rank_disagreement} shows the central result. Fidelity-only and QFC rankings disagree substantially: at 1000 shots IBM Marrakesh has the lowest mean TVD among the superconducting entries (0.0254) while IQM Sirius reaches the highest QFC (608.21), and QFC values span roughly 1.5--2 orders of magnitude, with IQM Sirius approximately $47\times$ the lowest-ranked entry, IBM Torino (retired). A third trend, that mean QFC increases from 100 to 1000 shots in every one of the 14 entries, is a property of the billing models rather than evidence that hardware quality improves with shot count; the next subsection separates the~two.

\subsection{Billing Model Determines Shot Scaling}
\label{sec:billing_regimes}

\noindent The rate at which QFC grows with $N$ is fixed almost entirely by the billing model. Substituting each cost expression into Equation~\eqref{eq:qfc_metric} with $\beta=\gamma=1$ makes this explicit: per-shot-only billing cancels the explicit shot term, giving $\mathrm{QFC}=e^{-\alpha D_{\mathrm{KL}}}/c_{\text{shot}}$, whose only $N$-dependence is through the fidelity term; task-plus-shot billing gives $e^{-\alpha D_{\mathrm{KL}}}N/(c_{\text{task}}+Nc_{\text{shot}})$, which rises as the task fee amortizes and saturates; and under runtime billing the billed duration is nearly constant over this shot range, so QFC grows about linearly. Provider-reported runtime is largely fixed overhead at this circuit size: IQM Sirius averages 1.34\,s at 50 shots and 1.73\,s at 1000, a $1.29\times$ increase for $20\times$ the work, and IBM Brisbane reports exactly 1\,s at 100, 500 and 1000 shots.

The measurements separate cleanly into these three bands (Figure~\ref{fig:qfc_all_qpus}). From 100 to 1000 shots the per-shot-only entry moves by $1.04\times$, the three task-plus-shot entries by $2.13$--$2.52\times$, and the ten runtime-billed entries by $5.96$--$34.31\times$. Consequently the ranking in Table~\ref{tab:qfc_1000_rank} is specific to a 1000-shot budget: OQC Toshiko Tokyo-1 ranks first at 50 and 100 shots, fifth of 14 at 500, and eleventh of 14 at 1000, without its fidelity changing appreciably. A cost-aware comparison is only meaningful at a stated shot budget, a requirement inherited from the billing models rather than a peculiarity of QFC.

Table~\ref{tab:qfc_1000_rank} gives the full 1000-shot ranking. The seven IBM devices span a $17\times$ range on their own, under one billing model at one per-second price, so the spread is not merely a vendor-pricing artifact. IBM Marrakesh illustrates the central disagreement directly: it records the best mean TVD in the superconducting set yet ranks fifth by QFC.

\begin{table}[!t]
\centering
\footnotesize
\caption{1000-shot QFC ranking across all measured QPU access-path entries (QFC standard deviation reported as population standard deviation; $n$ is the number of 1000-shot runs per entry).}
\label{tab:qfc_1000_rank}
\begin{tabularx}{\columnwidth}{@{}X c c c c c@{}}
\toprule
\textbf{QPU} & \textbf{Cloud} & \textbf{Mean QFC} & \textbf{Std. Dev.} & \textbf{Mean TVD} & \textbf{$n$} \\
\midrule
IQM Sirius            & IQM & 608.21 & 81.27  & 0.0502 & 5 \\
IQM Garnet            & IQM & 275.89 & 127.43 & 0.0436 & 5 \\
IQM Emerald           & AWS & 238.01 & 48.20  & 0.0360 & 3 \\
IQM Garnet            & AWS & 224.13 & 13.15  & 0.0490 & 3 \\
IBM Marrakesh         & IBM & 217.93 & 21.23  & 0.0254 & 5 \\
IBM Brisbane (ret.)   & IBM & 173.79 & 37.29  & 0.0576 & 5 \\
IBM Pittsburgh        & IBM & 156.69 & 12.44  & 0.0334 & 5 \\
IBM Kingston          & IBM & 149.03 & 59.89  & 0.0312 & 5 \\
IBM Fez               & IBM & 148.33 & 13.98  & 0.0332 & 5 \\
Rigetti Ankaa-3       & AWS & 138.55 & 14.13  & 0.0768 & 5 \\
OQC Toshiko Tokyo-1   & OQC & 120.47 & 10.02  & 0.0844 & 5 \\
IQM Emerald           & IQM &  73.82 & 14.76  & 0.0658 & 5 \\
IBM Sherbrooke (ret.) & IBM &  57.76 & 22.56  & 0.1044 & 5 \\
IBM Torino (ret.)     & IBM &  12.75 &  1.02  & 0.1300 & 5 \\
\bottomrule
\end{tabularx}
\end{table}

\subsection{Access-Path Observations}

For QPUs available through multiple access paths, cloud choice also matters. The two access paths for IQM Emerald produce different 1000-shot QFC under this pricing snapshot (AWS 238.01 vs.\ IQM Resonance 73.82), while for IQM Garnet the direction reverses (AWS 224.13 vs.\ IQM Resonance 275.89).

Because QFC is a product of a fidelity factor and a reciprocal cost factor, the gap can be attributed between the two. For IQM Emerald the $3.22\times$ advantage of the AWS path decomposes almost exactly into a $2.04\times$ fidelity factor (mean $D_{\mathrm{KL}}$ 0.817 vs.\ 1.528) and a $1.58\times$ cost factor (\$1.90 vs.\ \$3.00 per run): mostly a fidelity effect: the same physical QPU showed measurably different output quality across the two access paths. For IQM Garnet the decomposition does not close ($1.10\times$ predicted against $0.81\times$ observed) because the Resonance-path cost distribution is heavy-tailed. The AWS-path entries have only $3$ runs at 1000 shots, so we report both paths to show that the same hardware can land differently under different billing semantics, not to make an AWS-vs-IQM claim at this scale.

\subsection{A Trapped-Ion Point of Comparison}

\noindent The evaluated set is superconducting, but the same pipeline was run on one trapped-ion device, IonQ Aria-1 via AWS Braket, at $n=2$ runs per shot level. We report it outside Table~\ref{tab:qfc_1000_rank} because two runs cannot support a ranking claim. At 1000 shots Aria-1 records a mean TVD of 0.0245, the lowest fidelity error measured anywhere in this study, yet a QFC of 27.84, which would place it fourteenth of fifteen. The cause is entirely price: at \$0.03 per shot plus a \$0.30 task fee a 1000-shot run costs \$30.30, against \$0.60 for the top-ranked entry. The best-fidelity device measured here is thus roughly $22\times$ worse per dollar than the best-value device (Figure~\ref{fig:rank_disagreement}), indicating that the disagreement is not confined to one hardware modality.

\subsection{Sensitivity and Robustness}
\label{sec:sensitivity}

\noindent All results above use $\alpha=\beta=\gamma=1$. Because QFC is a weighted preference function, its absolute values necessarily move with those weights; the question that matters is whether the \emph{ranking} does. We recompute QFC over the saved per-run data under alternative parameterizations and report Kendall's $\tau$ against the published 1000-shot ordering (Table~\ref{tab:sensitivity}). No new hardware runs are needed, since the per-run cost of every execution is recoverable in closed form from the released counts and scores.

\begin{table}[!t]
\centering
\footnotesize
\caption{Ranking stability of the 1000-shot ordering under alternative sensitivity parameters and under an alternative IQM billing assumption. $\tau$ is Kendall's rank correlation against the default parameterization.}
\label{tab:sensitivity}
\begin{tabularx}{\columnwidth}{@{}l c X@{}}
\toprule
\textbf{Variant} & \textbf{$\tau$} & \textbf{Top-ranked entry} \\
\midrule
$\alpha{=}\beta{=}\gamma{=}1$ (default) & --- & IQM Sirius \\
$\gamma=0.5$ or $\gamma=0$ & $+1.000$ & IQM Sirius \\
$\beta=0.5$ & $+0.824$ & IQM Sirius \\
$\alpha=2$ & $+0.802$ & IQM Sirius \\
$\alpha=0.5$ or $\beta=2$ & $+0.714$ & IQM Sirius \\
$\alpha=5$ & $+0.560$ & IBM Marrakesh \\
\midrule
No IQM second-rounding & $+1.000$ & IQM Sirius \\
\bottomrule
\end{tabularx}
\end{table}

Three observations follow. First, $\gamma$ does not affect the ranking at a fixed shot count ($\tau=1.000$); it rescales every entry equally and matters only across shot budgets. Second, the ordering is stable under moderate reweighting ($\tau\ge0.71$ over a fourfold range of $\alpha$ and $\beta$), with IQM Sirius top-ranked in nine of ten parameterizations. Only strong fidelity weighting ($\alpha=5$) moves IBM Marrakesh to first, as expected once the KL term dominates cost: Marrakesh has the lowest mean KL, and here also the lowest mean TVD, among the entries. Third, billing IQM on exact rather than rounded seconds leaves the ordering identical and raises IQM Sirius to 708.50, so our rounding is conservative with respect to our own conclusion. The reported disagreement is a property of the measurements and billing models, not of the default weights.

\section{Related Work}
\label{sec:related_work}

Hardware-oriented evaluation has traditionally emphasized coherence times, gate errors, readout fidelity, and connectivity, with composite metrics such as Quantum Volume~\cite{cross_qv} and CLOPS~\cite{wack_clops} summarizing device and stack performance under NISQ conditions~\cite{preskill_nisq}. A parallel line evaluates algorithmic workloads directly, spanning circuit-level, application-level and hardware-agnostic scoring~\cite{li_qasmbench,tomesh_supermarq,donkers_qpack,lubinski_qedc,finzgar_quark,martiel_qscore}, and recent surveys argue that useful evaluation must span component, system, software, and cloud integration layers~\cite{lorenz_systematic_benchmarking}. Most of these frameworks treat monetary cost as secondary to performance; the community platform Metriq is a notable exception, reporting per-benchmark execution cost and pre-run cost estimation, though its composite score still aggregates performance subscores alone~\cite{metriq_cosentino}.

A separate body of work treats the quantum cloud as a resource-management problem. Qonductor orchestrates hybrid workloads across heterogeneous backends, estimating fidelity and runtime to balance user objectives against operator efficiency~\cite{giortamis_qonductor}; QSRA adapts CPU scheduling to QPUs to improve utilization and turnaround~\cite{lu_qsra}; and QuSplit raises aggregate fidelity and throughput by splitting jobs across backends of differing noise~\cite{li_qusplit}. Closest to our concerns, Szefer shows per-shot billing is itself exploitable: packing several circuits into one shot via mid-circuit reset can cut a workload's billed cost by up to an order of magnitude~\cite{szefer_reset_free}, underlining that a pricing model is a design surface, not merely a lookup table.

These two lines are complementary to ours rather than overlapping. Schedulers optimize \emph{predicted} fidelity, runtime, or utilization \emph{before} execution, and are evaluated on throughput and turnaround; QFC instead scores \emph{observed} executions \emph{after} the fact against the price actually billed, and is agnostic to how the job was placed. The two compose: QFC is directly usable as the objective a cost-aware scheduler maximizes. Where prior work reports fidelity and cost side by side, QFC folds measured fidelity and billed cost into a single score; to our knowledge it is the first to do so, giving users a view of where and how to best spend their budgets.

\section{Discussion}
\label{sec:discussion}

\noindent QFC is intended for three decisions: backend selection under a budget, computed from a short calibration run per candidate path; use as a scheduler objective, since cost is absent from current quantum cloud schedulers~\cite{giortamis_qonductor,lu_qsra,li_qusplit}, though that requires predicting rather than measuring the fidelity and runtime terms; and procurement monitoring, since an access path can be re-scored from saved counts whenever prices change. In all three the shot budget must be fixed before scores are compared (Section~\ref{sec:billing_regimes}).

Our testing budget bounds what the study can claim: the workload is a single Bell state, and QFC depends on the smoothing constant $\epsilon$, the parameters $\alpha,\beta,\gamma$, and a pricing snapshot. The $\epsilon$-dependence is not only in absolute values: varying $\epsilon$ from $10^{-6}$ to $10^{-15}$ shifts the 1000-shot ranking by Kendall $\tau$ of $0.67$--$0.89$ against the $\epsilon=10^{-12}$ default, comparable to the $\alpha$ and $\beta$ rows of Table~\ref{tab:sensitivity}, though IQM Sirius stays top throughout.

Three measurement-side caveats bound how finely the reported numbers should be read. First, the two runtime-billed providers report duration at different resolutions: IBM reports whole seconds, so its cost axis is quantized to \$1.60 and a typical two-second job carries up to $\pm50\%$ rounding error. Some shot-count movement of individual IBM entries is this quantization rather than device behavior; IBM Pittsburgh's large apparent gain from 100 to 1000 shots is driven by one 15\,s run at 100 shots against a uniform 2\,s at 1000. Second, runtime billing exposes users to infrastructure jitter unrelated to output quality: three IQM Garnet runs on the same day at the same shot count reported 1.69\,s, 5.84\,s and 6.47\,s and were billed \$1.00, \$3.00 and \$3.50, while their TVDs were 0.038, 0.038 and 0.043. That variance, not fidelity variance, dominates the standard deviation reported for the entry. Third, four IQM Garnet submissions failed with backend errors and are excluded from the 247 runs; one still reported 1.83\,s of billable runtime for no usable result. QFC scores only successful runs, so it does not charge a provider for its failure rate; a reliability-weighted variant is a reasonable extension.

Reported variances also understate exposure to recalibration: repeated runs were mostly back-to-back (twenty IQM Emerald runs span 55\,s, twenty IQM Sirius runs 45\,s), so they measure within-session rather than across-calibration variability. A QFC ranking is therefore attached to both a pricing snapshot and a calibration epoch. 

A natural follow-up work includes deeper circuits (GHZ, small QAOA/VQE ansatze), alternative scoring forms (different divergences or time-to-solution cost), and queue and wait-time cost. QFC rankings should also be re-computed as prices change each year.

\section{Conclusion}
\label{sec:conclusion}

This paper proposed Quantum Fidelity-per-Cost (QFC), a cost-aware score combining fidelity via KL divergence, shot count, and billed cost. It applied QFC to Bell-state execution across 14 cloud QPU access-path entries (12 physical QPUs, 247 runs), and considered reweighting of the QFC metric (Section~\ref{sec:sensitivity}). The results showed that the lowest-TVD device at 1000 shots was not the highest-QFC device, thus cost-aware ranking can differ from purely fidelity-based ranking. For algorithms that can tolerate certain levels of noise, our QFC offers a way for users to analyze which quantum computer, through which cloud provider, gives the best value for execution of their circuits. Prior ways of selecting quantum computer providers based on fidelity alone ignored important economic aspects and may cause users to spend excess money.

\section*{Acknowledgment}
The authors would like to thank IQM for IQM Resonance credits used in this work, and Oxford Quantum Circuits for Toshiko Tokyo-1 access. The authors used paid qBraid account for Amazon Braket access, while the IBM Quantum runs were done using IBM's pay-as-you-go service.

\clearpage
\bibliographystyle{IEEEtran}
\bibliography{refs}

\end{document}